# All-optical magnetization reversal via x-ray magnetic circular dichroism

Kihiro T. Yamada[1,*], Akira Izumi[2], Tetsuya Ikebuchi[3], Sumiyuki Okabe[2], Masaki Kubo[2], Ryusei Obata[2], Rei Kobayashi[4], Yuya Kubota[5], Takuo Ohkochi[5,6,7], Naomi Kawamura[6], Kotaro Higashi[6], Yoichi Shiota[3], Takahiro Moriyama[3,§], Teruo Ono[3,8], Iwao Matsuda[9], Tadashi Togashi[5,6], Yoshihito Tanaka[2,6,†], Motohiro Suzuki[4,5,6,‡]

[1]*Department of Physics, Institute of Science Tokyo, Meguro-ku, Tokyo, 152-8551, Japan.*

[2]*Graduate School of Science, University of Hyogo, Ako-gun, Hyogo, 678-1297, Japan.*

[3]*Institute for Chemical Research, Kyoto University, Uji, Kyoto, 611-0011, Japan.*

[4]*Program of Materials Science, School of Engineering, Kwansei Gakuin University, Sanda, Hyogo, 669-1330, Japan.*

[5]*RIKEN SPring-8 Center; Sayo-gun, Hyogo, 679-5148, Japan.*

[6]*Japan Synchrotron Radiation Research Institute, Sayo-gun, Hyogo, 679-5198, Japan.*

[7]*Laboratory of Advanced Science and Technology for Industry University of Hyogo, Ako-gun, Hyogo, 678-1205, Japan.*

[8]*Center for Spintronics Research Network, Osaka University, Toyonaka, Osaka, 560-8531, Japan.*

[9]*Institute for Solid State Physics, The University of Tokyo, Kashiwa, 277-8581, Japan.*

[§]*Department of Material Physics, Graduate School of Engineering, Nagoya University, Nagoya, 464-8603, Japan.*

*Contact author: yamada@phys.sci.isct.ac.jp

[†]Contact author: tanaka@sci.u-hyogo.ac.jp

[‡]Contact author: m-suzuki@kwansei.ac.jp.

**Light polarization is one of the most fundamental features, equivalent to energy and coherence. Magnetism changes light polarization, and vice versa. The irradiation of intense circularly polarized femtosecond pules to magnetic materials can alter the magnetic orders and elementary excitations, particularly in the visible to infrared spectral regions. Furthermore, the recent development of x-ray free-electron laser enables the element-specific trace of the ultrafast dynamics with high time and spatial resolution. However, the light helicity of x-ray photons has not yet been used to control order parameters in condensed matter materials, not limited to such magnetic phenomenon. Here, we demonstrate the deterministic magnetization reversal of a ferromagnetic Pt/Co/Pt multilayer solely by irradiating femtosecond pulses of circularly polarized hard x-rays. The observed all-optical magnetization switching depends on the helicity of incident x-ray pulses and is strongly resonant with the photon energy at the Pt**

**$L_3$ edge. These results originate in the x-ray magnetic circular dichroism of Pt, involving helicity-dependent excitation from the $2p_{3/2}$ core level to the exchange-split $5d$ valence states owing to the magnetic proximity effect with Co. These findings mark a new frontier for examining interactions between light and matter in the x-ray region.**

Magneto-optical effects, first discovered in the 19th century by Michael Faraday [1], are used for detecting the magnetism of condensed matter [2]. The polarization state of light passing through a magnetic material changes owing to electric dipole transitions depending on the light helicity and the spin and orbital states of the magnet [3]. Generally, in the infrared to ultraviolet spectral range, which corresponds to the photon energy of several electron volts, the magneto-optical effects originate from electric dipole transitions within the outermost electron shells. By further increasing the photon energy, magneto-optical effects can be resonantly and element-selectively induced when the photon energy matches the energy difference between the core level and the empty valence states of targeted magnetic atoms [4,5]. On the basis of sum rules, x-ray magnetic circular dichroism (XMCD) [6–10] is renowned for separately verifying the spin and orbital magnetic moments of a targeted element, particularly in the x-ray region [11,12]. In contrast, by exciting valence spin and orbital states of a magnetic material with intense circularly polarized laser pulses in the visible to near-infrared regions, all-optical manipulation of magnetization [13] has been achieved, such as the helicity-dependent excitation of magnons via the inverse Faraday effect [14] and all-optical helicity-dependent magnetization reversals [15,16]. However, it is unclear whether intense circularly polarized x-ray pulses can deterministically control magnetization via core-to-valence optical transitions.

The development of x-ray free-electron lasers (XFELs) [17–19] provide an opportunity to investigate all-optical helicity-dependent magnetization reversal via core-to-valence electric dipole transitions. XFEL facilities have been extensively used to conduct research in biology, chemistry, and physics because of the unique characteristics of x-rays, which exhibit high peak brilliance, high coherence, and ultrashort pulse width [20]. In particular, XFEL sources have recently enabled the exploration of highly excited states of condensed matter. This includes saturable absorption [21,22], multi-photon absorption [23,24], and ultrafast demagnetization [25,26] induced via the irradiation of intense XFEL pulses. However, although light helicity (that is, photon spin angular momentum) is one of the most distinct features of light, the helicity of XFEL pulses has not been fully used to intensively excite materials. Because XMCD arises from the difference in the transition probability of core-to-valence optical transitions based on the light helicity, XMCD would have crucial impacts on the intensively excited phenomena within the x-ray spectral range. In ferromagnets which are widely used in nonvolatile recording media, such as hard disc drive and magnetic random-access memory, the spin states are spontaneously split by the exchange interaction. Therefore, ferromagnets can be an ideal platform because x-ray photon spins on the electron spins can be transcribed via XMCD and semi-permanently stored in macroscopic spin configurations.

In this Article, we demonstrate the all-optical helicity-dependent magnetization reversal of a ferromagnetic Pt/Co/Pt multilayer film via XMCD at the $L_3$ edge of Pt in the hard x-ray region. Pt/Co/Pt multilayers are widely used for all-optical helicity-dependent magnetization switching [16,27–31] in the visible to near-infrared regions. Figure 1(a) illustrates the research concept. Intense circularly polarized XFEL pulses resonantly excite electrons in the $2p_{3/2}$ core levels to the empty $5d$ valence states of Pt. Here, the probability of the core-to-valence electric dipole transition depends on the light helicity and the spontaneous spin polarization of Pt induced by the ferromagnetic proximity effect [9,10]. Consequently, the irradiation of intense circularly polarized hard XFEL pulses to the perpendicularly magnetized Pt/Co/Pt multilayer switches the macroscopic magnetization distribution in a helicity-dependent manner.

We prepared a multilayer consisting of Ta (1 nm)/[Pt (1 nm)/Co (0.5 nm)]$_3$/Pt (1 nm)/Ta (2 nm) by DC sputtering on a synthetic quartz glass substrate with a thickness of 0.5 mm. The XMCD of the prepared Pt/Co/Pt multilayer was characterized at BL39XU of SPring-8 (see Supplemental Note 1 for more details of the XMCD characterization [32]). Figure 1(b) indicates that the XMCD peak is approximately 8% of the edge jump of x-ray absorption spectra at the Pt $L_3$ edge. The XMCD amplitude, as a function of an external out-of-plane magnetic field, exhibits a square-hysteresis response [Fig. 1(c)], indicating significant uniaxial perpendicular magnetic anisotropy. We observed the multi-pulse all-optical helicity-dependent switching of the Pt/Co/Pt multilayer using 60-fs-wide near-infrared laser pulses with photon energy ($E$) of 1.55 eV [see Fig. S2 in the Supplemental Note 3 [32]].

We used BL 3 of SACLA [19,33], in which intense monochromatized and circularly polarized XFEL pulses are available in the hard x-ray region. A uniformly magnetized multilayer was exposed to XFEL pulses with a degree of circular polarization higher than 90 percent [34,35], at a duration of approximately 7 fs FWHM [36–38] and a repetition rate of 30 Hz. X-ray refractive lenses were used to focus the x-ray beam to an elliptical spot with dimensions of 20 μm × 16 μm, such that the x-ray fluence was increased to 72 $mJ/cm^2$ (see Supplemental Notes 4–7 [32]). During the irradiation of XFEL pulses, the sample position was continuously translated within the film plane using motorized stages. This prototypical scanning approach [15,16,28–31] allows helicity-biased magnetic domains to remain visible even when the helicity-dependent effect is indistinguishable under fixed irradiation. The switching rate of multi-pulse all-optical magnetization switching increases with decreasing scan speed in the near-infrared regime [27,29]. Therefore, the scan speed was set to 0.5 μm/s, which is the slowest possible within the limited beam time. In addition, the slow scan speed helped minimize the effects of pulse-energy and beam-position fluctuations. Magnetic domains were visualized using a polarization microscope with a halogen light source. Magnetization contrasts were obtained from microscopic images by analyzing the magneto-optical Faraday rotation angles ($\theta_F$) of linearly polarized white light passing through the magnetic multilayer. The weak decay of the hard XFEL pulses in the air facilitated the construction of the magneto-optical-imaging setup. See Supplemental Note 2 [32] for details of the measurement configurations.

First, the magnetic multilayer was excited using circularly polarized XFEL pulses at $E$ = 11.569 keV, at which the XMCD effect was maximized around the white line of Pt $L_3$ XAS. Figure 2(a) shows magneto-optical images after the irradiation of right-handed circularly ($\sigma^+$), left-handed circularly ($\sigma^-$), and randomly ($r$) polarized XFEL pulses. On the exposure tracks of XFEL pulses, the spatial distributions of magnetic contrast (that is, either red or blue) changed slightly, depending on the light helicity. The enhancement at the domain boundaries arises from light scattering by magnetic domain walls, not from intrinsic magnetization signals.

To quantify the helicity-dependence of magnetic contrasts, we calculated histograms of $\theta_F$ [Fig. 2(b)] by extracting the $\theta_F$ values in the irradiated areas. We excluded counts with $|\theta_F| > 0.7°$ originating from light scattering by defects in the multilayer. Counts corresponding to unirradiated areas owing to accidental x-ray beam abortions were also excluded. The left and right panels in Fig. 2(b) indicate clearly biased distributions: the mode of the histogram shifts toward the positive (negative) direction for the $\sigma^+$ ($\sigma^-$) helicity of the x-ray pulses regardless of the initial magnetization direction. In the middle panel of Fig. 2(b), for the $r$ polarization, no significant change in the histogram exists for either initial magnetization direction. The significant helicity dependence of the arithmetic means $\mu_{\theta_F}$ of Faraday rotation in Fig. 2(c) indicates that the observed all-optical magnetization switching genuinely relies on the helicity of circularly polarized XFEL pulses.

The observed helicity-dependent component is approximately $10\,\%$ of the Faraday rotation since the driving force is relatively weak. The magnetization reversal process thus requires thermal fluctuation to assist the nucleation and growth of flipped magnetic domains. These characteristics suggest that the magnetization reversal proceeds via a multi-pulse process, similar to conventional all-optical helicity-dependent magnetization switching in the visible-to-near-infrared regime [27]. Accordingly, in the present experiments using circularly polarized hard x-ray pulses, the magnetization reversal is expected to occur through helicity-dependent domain wall motion followed by the helicity-independent formation of reversed magnetic domains. We note, however, that the similarity in the magnetization switching process does not necessarily imply an identical driving mechanism.

In addition, the photon-energy dependence of the all-optical magnetization reversal was examined by scanning the photon energy around the Pt $L_3$ edge. The magneto-optical images in Fig. 3 show the magnetic domains after the irradiation of XFEL pulses at $E$ = 11.549, 11.566, 11.569, 11.572, and 11.589 keV. To precisely verify the helicity dependence of the magnetization reversals, we defined $\Delta\mu_{\theta_F} = \mu_{\theta_F}[(M^\uparrow, \sigma^-) - (M^\uparrow, \sigma^+) + (M^\downarrow, \sigma^-) - (M^\downarrow, \sigma^+)]/4$ to extract the helicity-dependent changes and eliminate the magnetization-dependent contribution. $\Delta\mu_{\theta_F}$ is plotted as a function of $E$ [Fig. 4] and is consistent with the XMCD spectrum at the Pt $L_3$ resonance. Therefore, the results show that the helicity-dependent excitation of core-to-valence electric dipole transitions via XMCD causes the helicity-dependent magnetization reversal of the Pt/Co/Pt multilayer via irradiation of circularly polarized XFEL pulses. The magneto-optical Faraday effect with the white light probes the total magnetization of the valence band consisting of Co 3*d* and Pt 5*d* orbits. Thus, the results demonstrate that magnetization can be reversed by selectively exciting the core-level electrons of Pt, whose magnetization emerges from the interfacial proximity effect dominated by inherently ferromagnetic Co atoms. Because the time scale of demagnetization of ferromagnetic metals is generally hundreds of femtoseconds [39], which is significantly longer than the XFEL pulse duration of 7 fs in this experiment, the exchange spin splitting in the Pt 5*d* sub-bands should be maintained during the excitation of core electrons occurring in the initial step. Hence, demagnetization should be associated with significant contributions of spin angular momenta transfer from Pt atoms to Co atoms [40,41]. Further experimental studies using the x-ray pump-and-probe technique [26] and theoretical approach [42] will help reveal the x-ray induced magnetization dynamics in an element-specific manner, addressing the origin of the reversed master (Co) and servant (Pt) relationship.

The excellent agreement of $\Delta\mu_{\theta_\mathrm{F}}$ with XMCD [Fig. 4] indicates that the observed magnetization reversal requires resonant absorption of circularly polarized x-ray photons. Therefore, the observed magnetization reversals are not caused by the real part of the resonant magneto-optical effect, that is, the inverse Faraday effect, of which the spectral dependence is given by the derivative of that of the imaginary part [4], based on the Kramers–Kronig relations. For the $L_3$ edge of nonmagnetic Pt, owing to spin-orbit splitting and the orbital selection rule, in a one-electron picture, a $\sigma^+$ x-ray photon with a spin angular momentum of $+\hbar$ excites an up-spin electron in the $2p_{3/2}$ core levels to the 5*d* valence up-spin sub-bands at a higher rate than when exciting a core down-spin electron to the valence down-spin sub-bands. This unequal probability of electric dipole transition is boosted particularly when the proximity-induced spin polarization of 5*d* electron (hole) bands is antiparallel (parallel) to the photon spin angular momentum (that is, the x-ray propagation direction). Consequently, nonequilibrium down-spin polarizations are formed in the 5*d* electron sub-bands, particularly at the $L_3$ edge. Although the photon flux per unit energy decreases in the x-ray range compared to in the visible spectral range, resonant excitation with circularly polarized x-ray pulses can promote the accumulation of spin-polarized electrons immediately above the Fermi level of the 5*d* sub-bands, resulting in a nonequilibrium spin-polarized state. Because the time scale of magnetization reversal dynamics is significantly longer than the duration of the nonequilibrium spin polarization regulated by the core-hole lifetime of 0.8 fs [43], the nonequilibrium spin polarization in the 5*d* sub-bands can affect the ultrafast demagnetization and subsequent remagnetization processes of the entire magnetic system via ferromagnetic *d*-*d* exchange coupling at the interface. Therefore, rather than magnetization switching by a single XFEL pulse, the accumulative effects of multiple $\sigma^{+(-)}$ x-ray pulses may gradually change the spatial magnetization distribution from the down (up) to up (down) direction. Moreover, the spin temperatures may differ between up-spin and down-spin sub-bands, resulting in helicity-dependent domain growth switching similar to the cases of all-optical magnetization reversal in the near-infrared to visible regions [30,31]. In the hard x-ray range, high-energy electrons created through processes, such as the Auger process, cause scattering all over the Brillouin zone and induce an exchange in the spin angular momentum and energy penetrating the layers, which might effectively generate the spin temperature difference. XMCD can give rise to multiple microscopic processes that may lead to helicity-dependent magnetization reversal. The observation of all-optical magnetization reversal via XMCD, together with further identification of the dominant process, is expected to stimulate future research into how circularly polarized high-energy photons interact with condensed matter materials.

We have demonstrated the all-optical magnetization reversal of the Pt/Co/Pt multilayer by circularly polarized hard XFEL pulses via the XMCD from helicity-dependent core-to-valence electric dipole transitions at the Pt $L_3$ edge. In future studies, x-ray-induced magnetization reversal caused by exciting the *L* and *M* edges of Co should be investigated for comparison. In particular, the XMCD at the $L_3$ edge of Co is an order of magnitude higher than that at the $L_3$ edge of Pt [7–10]; however, the photon energy is a lower order of magnitude. Therefore, the comparison of the results obtained in this study for the $L_3$ edge of Co should unravel not only the requirements for the deterministic magnetization reversal but also the origin of the observed reversed master–servant relationship. By selecting the absorption edge,

XFEL pulses can intensively excite the specific element and orbital states of condensed matter materials, not limited to magnets, such as superconducting [44] and chiral [45] materials. In addition to the diverse future directions of controlling physical properties using XFEL, the present observation of the intensively excited helicity-dependent phenomena in the x-ray region is valuable for further studies on light–matter interactions in condensed matter.

*Acknowledgments*— We acknowledge SACLA's workers for their assistance in conducting the XFEL measurements. We also thank A. V. Kimel for critically reading the manuscript. The XFEL experiment was performed at SACLA with the approval of JASRI and the Program Review Committee (Nos. 2021A8025, 2018A8020, 2019B8010, and 2020A8014). This research was partially supported by the MEXT/JSPS KAKENHI (Grant Nos. 17H02823, 20H00349, 19H04397, 22K14588, 20K22327, 24K00938, 23K04624, 24K21043, 21H05012, 21H05012, 20H05665, 21H04562, and 20K15161), the Inamori Research Grants, the Science Research Promotion Fund, the Promotion and Mutual Aid Corporation for Private Schools of Japan, the ASUNARO Grant, Institute of Science Tokyo, the Collaborative Research Program of the Institute for Chemical Research, Kyoto University, and the National Institutes of Natural Sciences (NINS) Frontier Photonic Sciences Project (Grant Nos. 01212405 and 01212307).

*Data availability* — The data that support the findings of this article are not publicly available. All data supporting the findings within this paper are available from the corresponding author upon reasonable request.

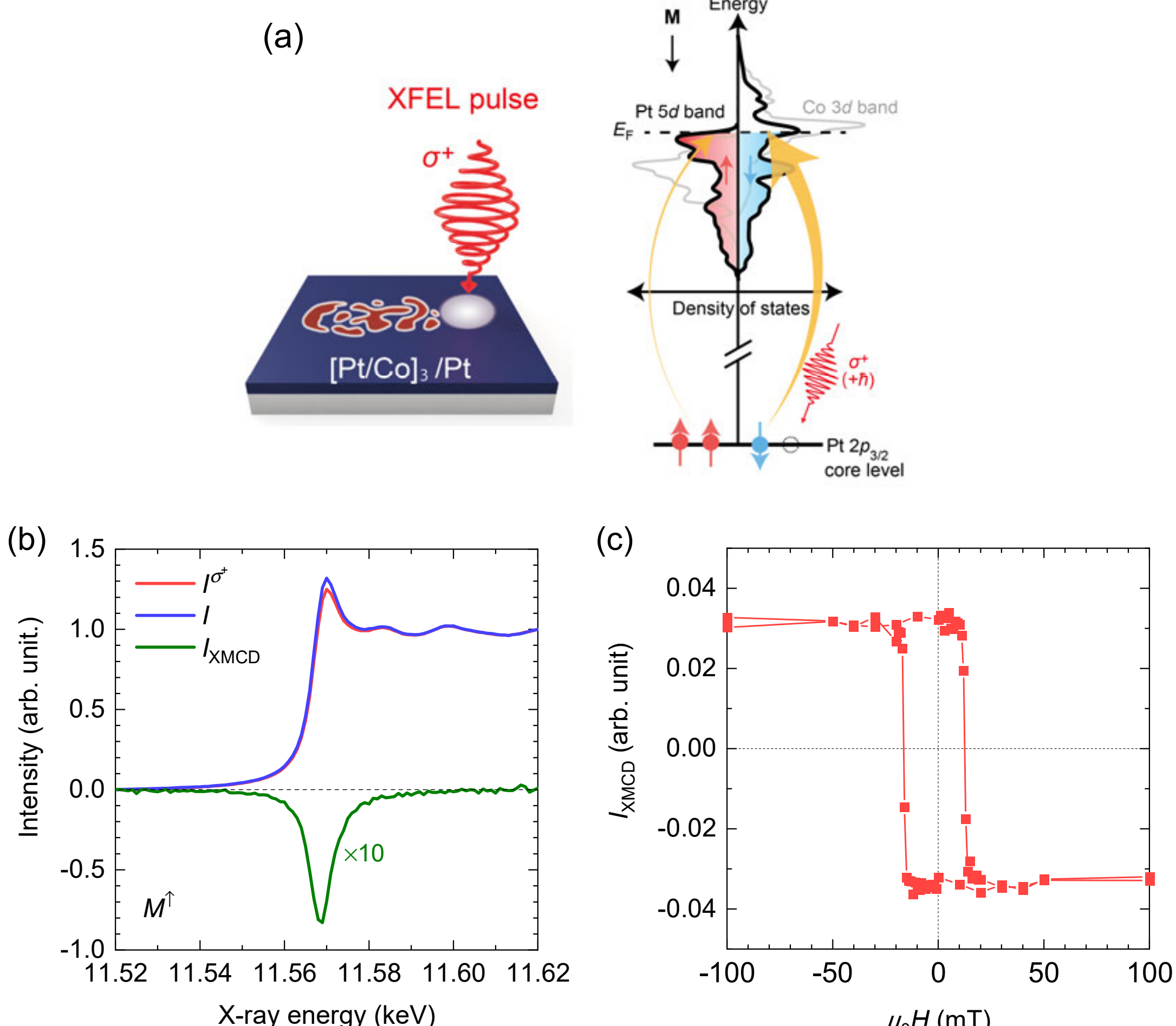

FIG. 1. Concept of magnetization reversal via x-ray magnetic circular dichroism (XMCD) in a Pt/Co/Pt multilayer and the magnetic properties. (a) (Left) Illustration of magnetization reversal by irradiation of circularly polarized x-ray free electron laser (XFEL) pulses. (Right) Diagram of the helicity-dependent core-to-valence electric dipole transitions of Pt at the $L_3$ absorption edge. The $5d$ electronic band structure of Pt within the proximity of Co exhibits spin splitting induced by the ferromagnetic exchange interaction between the $3d$ orbitals of Co and the $5d$ orbitals of Pt. In the coordinates, the right-handed ($\sigma^+$) and left-handed ($\sigma^-$) circularly polarized x-ray beams carry the spin angular momenta of $+\hbar$ and $-\hbar$ in the propagation direction of light, respectively. When exciting the down-magnetized multilayer with $\sigma^+$ XFEL pulses, down-electron spins in the Pt $2p_{3/2}$ core level preferentially transit to the down-spin $5d$ valence states. (b) XMCD at the $L_3$ absorption edge of Pt. Here, we define the XMCD intensity ($I_{\mathrm{XMCD}}$) as a difference between the x-ray florescence yields for $\sigma^+$ and $\sigma^-$ x-ray beams: $I_{\mathrm{XMCD}} = I^{\sigma^+} - I^{\sigma^-}$. (c) XMCD as a function of a perpendicular magnetic field. The photon energy is fixed at the $L_3$ white line, and $E = 11.569$ keV.

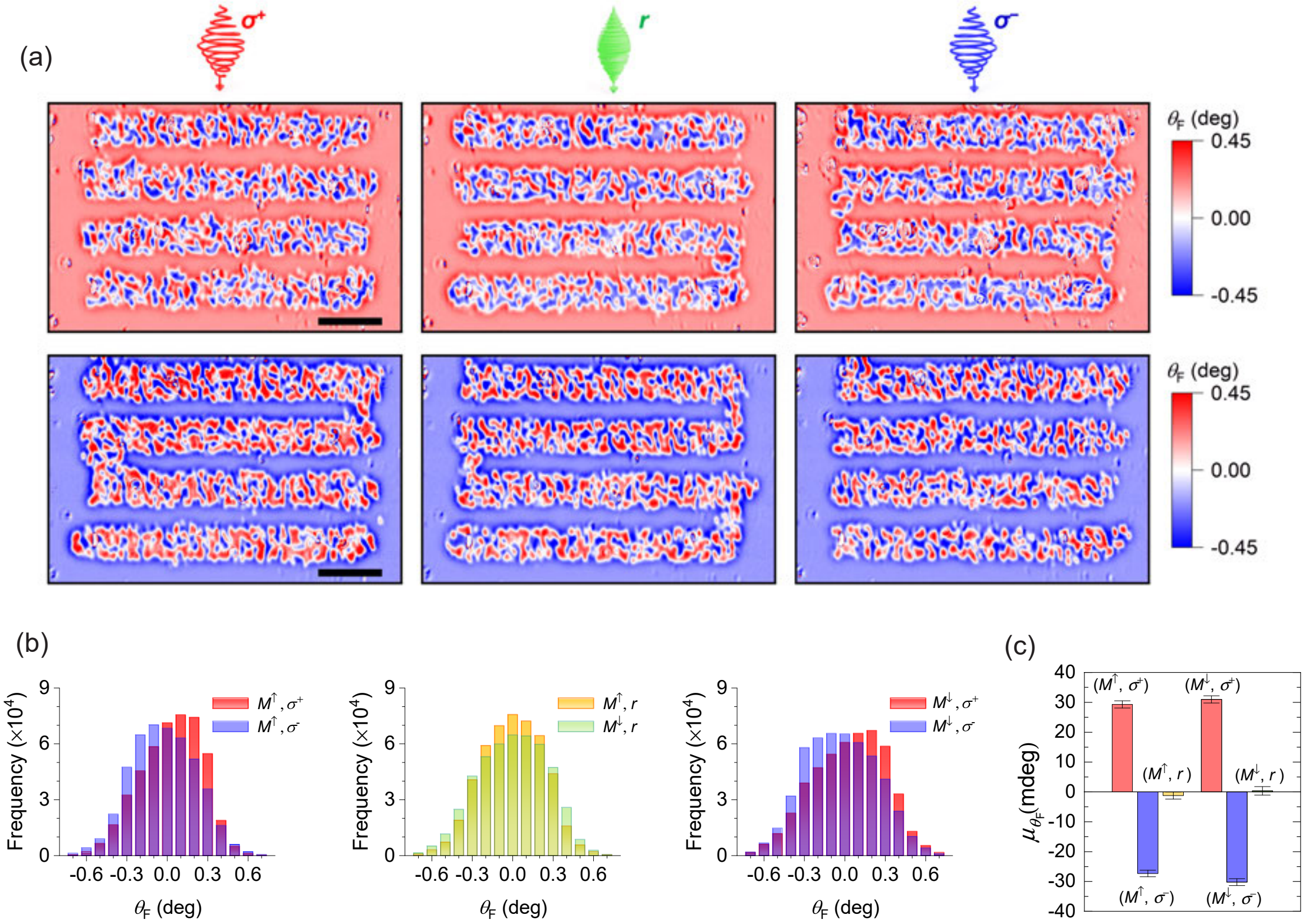

FIG. 2. All-optical helicity dependent magnetization reversal at the $L_3$ white line of Pt. (a) Faraday-rotation ($\theta_F$) maps after scanning hard XFEL pulses with circular and random polarizations. The photon energy of 11.569 keV corresponds to the XMCD peak of the Pt $L_3$ XAS. The red and blue regions represent up-magnetized ($M^{\uparrow}$) and down-magnetized ($M^{\downarrow}$) domains, respectively. The scale bars are 20 μm. (b) Histograms showing the values of $\theta_F$ inside the exposed areas. Each histogram sums 10 linear scans. See the other line scans in Figs. S5 [32]. The figure legends indicate the initial magnetization direction and x-ray polarization. (c) Arithmetic means of Faraday rotation $\mu_{\theta_F}$. The error bars correspond to the standard errors.

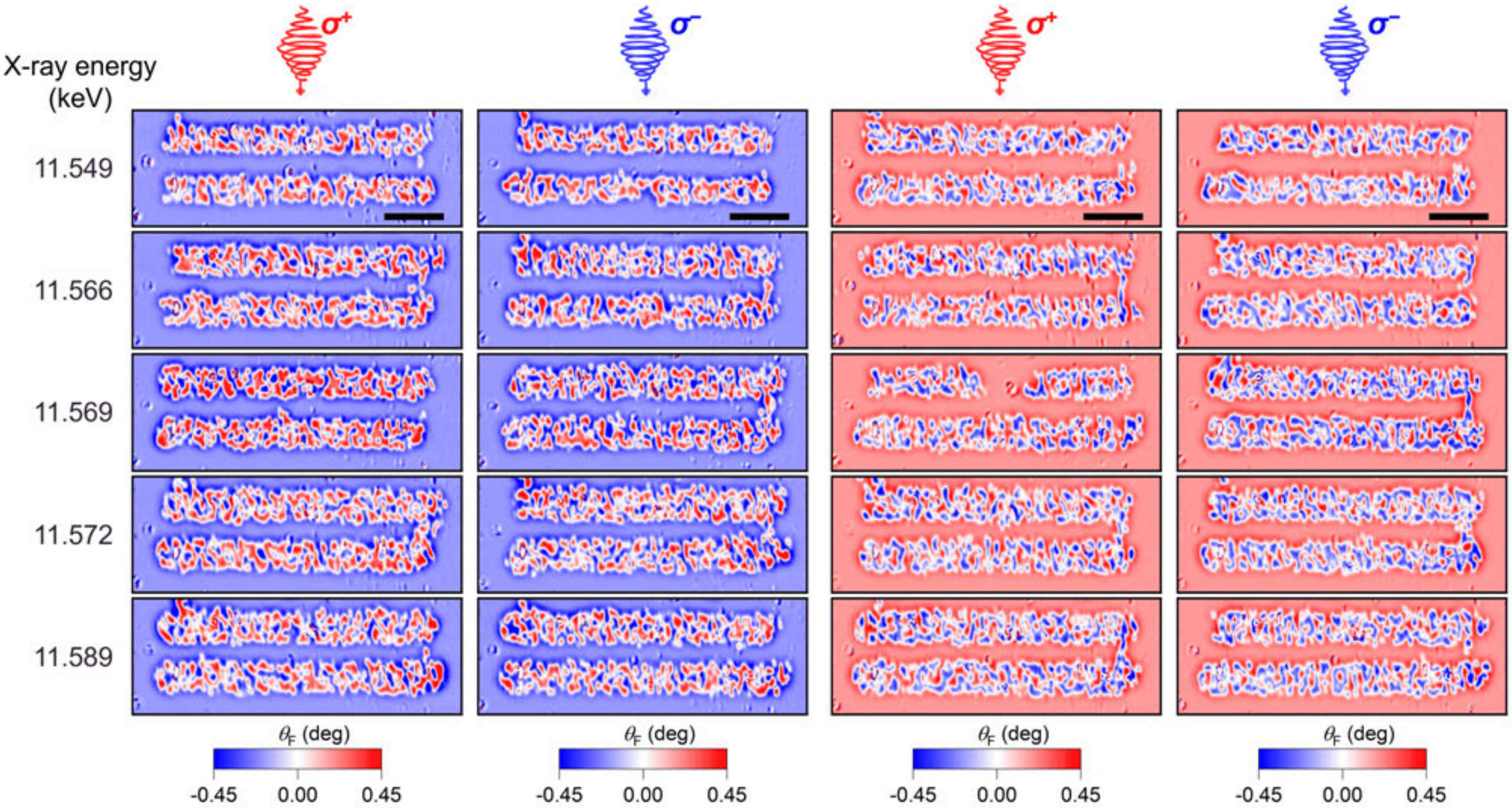

FIG. 3. Faraday-rotation ($\theta_F$) images after the irradiation of circularly polarized hard XFEL pulses as a function of photon energy around the Pt $L_3$ edge. The scale bars represent 20 μm. See the other line scans in Fig. S6 [32]..

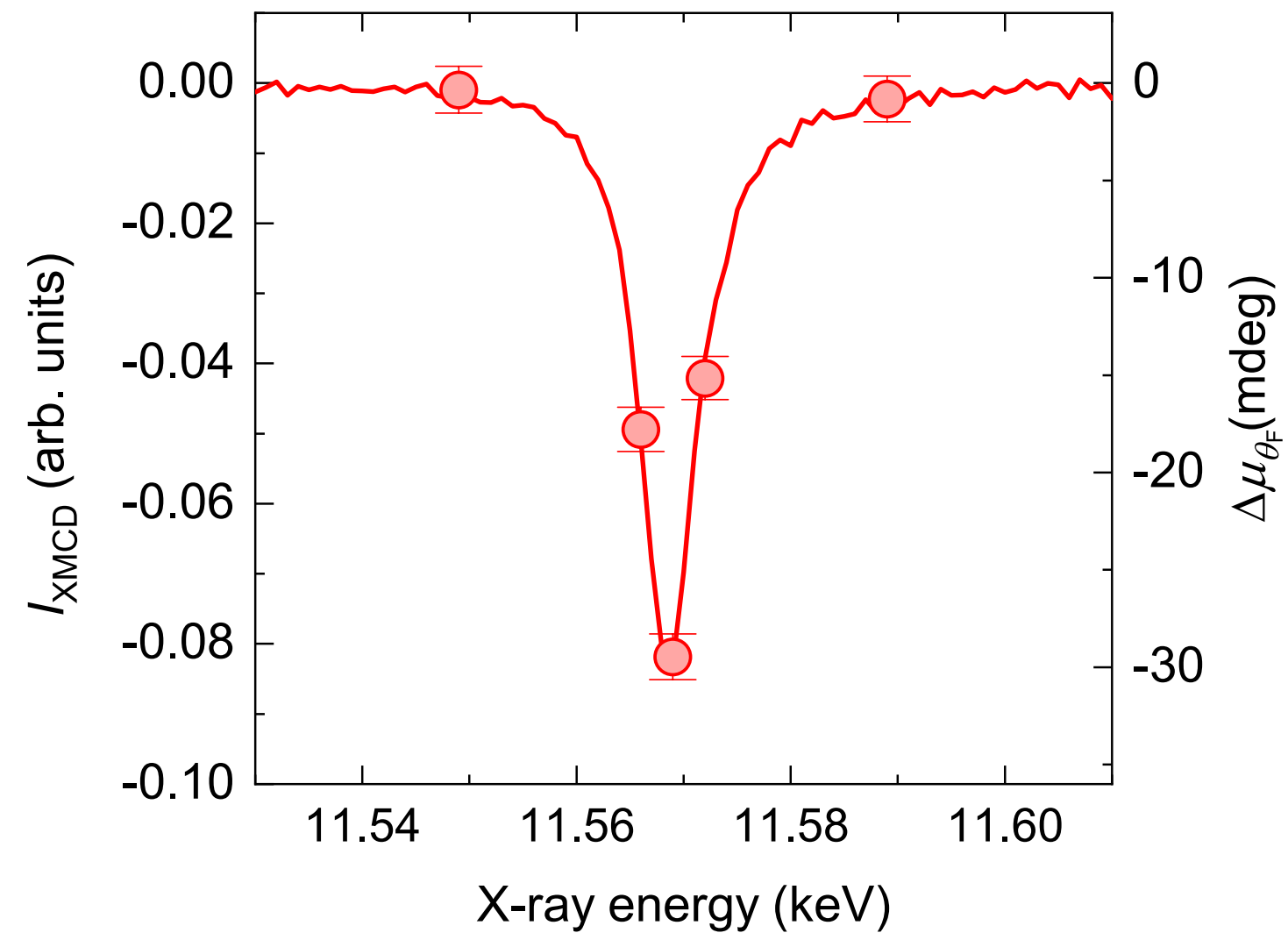

FIG. 4. Photon energy dependence of a difference of arithmetic means of Faraday rotation, $\Delta\mu_{\theta_F}$, between the use of $\sigma^+$ and $\sigma^-$ pulses. The red circles indicate $\Delta\mu_{\theta_F}$ values averaged for five scans at $E$ = 11.549, 11.566, 11.572, and 11.589 keV, and for 10 scans at 11.569 keV. For comparison, the XMCD spectrum in Fig. 1(b) is plotted with a solid line. See the $\theta_F$ histograms in Fig. S7 and calculated $\mu_{\theta_F}$ values at each helicity, magnetization, and energy condition in Table S4 [32].

## Supplemental Material

## All-optical magnetization reversal via x-ray magnetic circular dichroism

Kihiro T. Yamada[1,*], Akira Izumi[2], Tetsuya Ikebuchi[3], Sumiyuki Okabe[2], Masaki Kubo[2], Ryusei Obata[2], Rei Kobayashi[4], Yuya Kubota[5], Takuo Ohkochi[5,6,7], Naomi Kawamura[6], Kotaro Higashi[6], Yoichi Shiota[3], Takahiro Moriyama[3,§], Teruo Ono[3,8], Iwao Matsuda[9], Tadashi Togashi[5,6], Yoshihito Tanaka[2,6,†], Motohiro Suzuki[4,5,6,‡]

[1]*Department of Physics, Institute of Science Tokyo, Meguro-ku, Tokyo, 152-8551, Japan.*

[2]*Graduate School of Science, University of Hyogo, Ako-gun, Hyogo, 678-1297, Japan.*

[3]*Institute for Chemical Research, Kyoto University, Uji, Kyoto, 611-0011, Japan.*

[4]*Program of Materials Science, School of Engineering, Kwansei Gakuin University, Sanda, Hyogo, 669-1330, Japan.*

[5]*RIKEN SPring-8 Center; Sayo-gun, Hyogo, 679-5148, Japan.*

[6]*Japan Synchrotron Radiation Research Institute, Sayo-gun, Hyogo, 679-5198, Japan.*

[7]*Laboratory of Advanced Science and Technology for Industry University of Hyogo, Ako-gun, Hyogo, 678-1205, Japan.*

[8]*Center for Spintronics Research Network, Osaka University, Toyonaka, Osaka, 560-8531, Japan.*

[9]*Institute for Solid State Physics, The University of Tokyo, Kashiwa, 277-8581, Japan.*

[§]*Department of Material Physics, Graduate School of Engineering, Nagoya University, Nagoya, 464-8603, Japan.*

*Contact author: yamada@phys.sci.isct.ac.jp

[†]Contact author: tanaka@sci.u-hyogo.ac.jp

[‡]Contact author: m-suzuki@kwansei.ac.jp.

## Supplemental Note

### Supplemental Note 1: Fabrication and characterization of samples using x-ray absorption spectroscopy

We fabricated a magnetic multilayer consisting of Ta(1)[Pt(1)/Co(0.5)]$_3$/Pt(1)/Ta(2) (unit: nm) on a synthetic quartz glass substrate by DC sputtering [S1] at room temperature. The x-ray absorption spectroscopy (XAS) and x-ray magnetic circular dichroism (XMCD) spectra were measured at BL39XU of the SPring-8 synchrotron radiation facility [S2,S3]. Circularly polarized monochromatic x-ray beams with a photon energy ($E$) around the Pt $L_3$ edge (~11.6 keV) were incident on the multilayer. The x-ray fluorescence yields of the Pt $L_\alpha$ line were detected using a four-element silicon drift detector. A magnetic field of 1.0 T was applied using an electromagnet in the direction parallel to the x-ray beam, which was perpendicular to the film plane. According to the standard definition of the XMCD sign [S4], the following definitions [S5] of the handedness of circularly polarized light were adopted: the electric field vector of right-handed circularly polarized light projected onto a screen rotates clockwise when the electric field vector is observed with our back against the light source. On the basis of this definition, right-handed ($\sigma^+$) and left-handed ($\sigma^-$) circularly polarized x-ray beams carry spin angular momenta of $+\hbar$ and $-\hbar$, which are parallel and antiparallel to the propagation direction of the x-ray beam, respectively. The directions of up ($M^\uparrow$) and down ($M^\downarrow$) magnetizations are antiparallel and parallel to the x-ray propagation direction, respectively. Hence, the XAS intensity is determined by

$$I_{\text{XAS}} = \left(I^{\sigma^+} + I^{\sigma^-}\right)/2, \qquad (1)$$

where $I^{\sigma^{+(-)}}$ is the florescence yield for the $\sigma^{+(-)}$ helicity, which is normalized to the intensity of the incident x-ray beam. The XMCD intensity is defined as

$$I_{\text{XMCD}} = I^{\sigma^+} - I^{\sigma^-}. \qquad (2)$$

The XAS and XMCD spectra were further normalized by the XAS edge jump from $E$ = 11.620 keV to 11.520 keV (Fig. 1b). An element-specific magnetic hysteresis loop was obtained by monitoring $I_{\text{XMCD}}$ at the XMCD peak energy ($E$ = 11.569 keV) as a function of the out-of-plane magnetic field. See Ref. [S3] for further details of the XMCD measurement configuration. The measured XMCD spectrum with a negative sign at the Pt $L_3$ edge and the square hysteresis loop (Fig. 1b and 1c) indicate that the induced magnetic moments in the Pt layers are aligned perpendicular to the film and ferromagnetically coupled with the magnetic moments in the Co layers.

### Supplemental Note 2: Experimental setup for XFEL measurements

We analyzed the magnetization reversal of the Pt/Co/Pt multilayer induced by intense circularly polarized hard x-ray pulses at BL3 of the SPring-8 Angstrom Compact free-electron LAser (SACLA) [S6]. A schematic of the experimental configuration combining the XFEL with a static magneto-optical imaging setup is shown in Fig. S1 in the Supplementary Information. Coherent linearly polarized x-ray pulses radiated from the undulator through the self-amplified spontaneous emission process. The x-ray pulse outputs from the undulator at a duration of ~7 fs [S7–S9] were monochromatized using a Si 111 double-crystal monochromator with a band width of approximately 1 eV. The circular polarization was controlled using an x-ray phase retarder made of a 1.5-mm-thick diamond (100) crystal [S10][S10]. The photon polarization was switched between the $\sigma^{+}$ and $\sigma^{-}$ helicities by adjusting the crystal angle to a positive or negative angular offset from the Bragg condition of the 220 symmetric Laue reflection. A random polarization state was obtained at the exact Bragg angle. Next, an array of compound refractive lenses was used to focus the XFEL pulses on the magnetic multilayer placed in the air.
To visualize the magnetic states of the multilayer, we constructed a static magneto-optical microscope with a halogen fiber optic illuminator. The light source spectrum included the near-infrared to visible regions, with photon energies ranging between 0.7 and 3 eV. The fiber output was linearly polarized using a wire grid polarizer and collimated using an aspheric lens. The linearly polarized light was focused onto the multilayer using a plano-convex lens with a focal length of 150 mm. When the light passed through the magnetized multilayer, the polarization direction rotated owing to the magneto-optical Faraday effect, which depended on the out-of-plane magnetization component. Next, the transmitted light was collected using an objective lens with a magnification of ×50. A wire grid analyzer, positioned behind the objective lens, converted the Faraday rotation angle into intensity change, enabling the capture of microscopic magnetic images using a complementary metal-oxide semiconductor camera and an imaging lens. The multilayer sample was uniformly magnetized using an electromagnet before being exposed to XFEL pulses. Subsequently, the magnetic multilayer was excited at a zero-magnetic field because a small stray field could affect the final magnetic states. To minimize the influence of the remanent field from the poles of the electromagnet, we placed the multilayer on a nonmagnetic duralumin plate and adjusted the electromagnet power supply with a low offset voltage to cancel the leakage magnetic field. These measures reduced the magnetic flux density to lower than 0.03 mT at the sample position.

**Supplemental Note 3: All-optical magnetization reversal by circularly polarized near-infrared (NIR) pulses**

For comparison with the x-ray experimental results, we investigated all-optical magnetization switching of the same Pt/Co/Pt multilayer by circularly polarized near-infrared (NIR) pulses. First, the magnetic circular dichroism (MCD) of the multilayer was verified at a photon energy ($E$) of 1.55 eV. The MCD is defined as

$$\delta_{\mathrm{MCD}} = \frac{P_{\mathrm{a}}^{\sigma^{+}} - P_{\mathrm{a}}^{\sigma^{-}}}{(P_{\mathrm{a}}^{\sigma^{+}} + P_{\mathrm{a}}^{\sigma^{-}})/2}, \qquad \text{(S1)}$$

where $P_{\mathrm{a}}^{\sigma^{+(-)}}$ is the averaged absorbed power for $\sigma^{+(-)}$ helicity by the magnetic multilayer. The absorbed power was indirectly measured by subtracting the sum of the averaged reflected power ($P_{\mathrm{r}}$) and transmitted power ($P_{\mathrm{t}}$) from the averaged incident power ($P_{\mathrm{i}}$). To cancel the small helicity dependence of the power meter and the slight variation of the incident power by the rotation of a quarter-wave plate, we calculated the normalized power ratio: $P_{\mathrm{a}} / P_{\mathrm{i}} = 1 - (P_{\mathrm{r}} + P_{\mathrm{t}})/P_{\mathrm{i}}$. Table S1 lists the values for the irradiation of $\sigma^{+}$ or $\sigma^{-}$ light to a uniformly up-magnetized or down-magnetized film. The incident angle was 10° from the film normal. We used a collimated beam from a Ti:sapphire laser with a regenerative amplifier. The collimated beam had a power of 9.50 mW and a diameter of approximately 4 mm. The resulting fluence ($F$) of 0.08 mJ/cm$^{2}$ was significantly lower than the threshold of all-optical magnetization switching of the magnetic multilayer, and the static MCD signal was adequately determined. The measured value of NIR

MCD was $\delta_{\mathrm{MCD}}$ = 0.007±0.004, which is one order of magnitude lower than the XMCD at the Pt $L_3$ edge.

For all-optical magnetization switching, we also employed the Ti:sapphire laser with a regenerative amplifier. The repetition rate, pulse duration, and central wavelength were 1 kHz, 60 fs, and 800 nm, respectively. After the circular polarization of optical pulses was controlled using an achromatic quarter waveplate, we focused the pulses perpendicular to the sample surface using a lens. The $1/e^2$ radius was 71.1 μm. The magnetic multilayer was mounted on a motorized linear stage to scan the laser spot. Magneto-optical Kerr microscopy was applied in a reflection geometry to observe the magnetic domains. Linearly polarized incoherent light from a red light-emitting diode was incident on the multilayer from the substrate side. When the magnetic multilayer reflected the light, the linear polarization direction rotated owing to the magneto-optical Kerr effect. The reflected light was collimated by an objective lens with a magnification of 10 and analyzed using a wire-grid analyzer. The analyzed light formed a magneto-optical image on the light-receiving surface of a CMOS camera by an imaging lens.

We scanned the spot of circularly polarized NIR pulses on the surface of the magnetic multilayer at a speed of 18.4 μm/s and $F$ = 0.50 mJ/cm$^2$. Figure S2 shows the spatial maps of magneto-optical Kerr rotation ($\theta_{\mathrm{K}}$) after exciting up-magnetized ($M^{\uparrow}$) and down-magnetized ($M^{\downarrow}$) multilayer by $\sigma^+$ and $\sigma^-$ NIR pulses. The irradiation of $\sigma^{+\,(-)}$ pulses stabilized the $M^{\downarrow}$ ($M^{\uparrow}$) state. The sense is opposite to the case of magnetization reversal using circularly polarized XFEL pulses because the sign of NIR MCD is opposite to that of XMCD. We defined the helicity dependence of $\theta_{\mathrm{K}}$ as $\Delta\mu_{\theta_{\mathrm{K}}} = \mu_{\theta_{\mathrm{K}}}[(M^{\uparrow}, \sigma^-) - (M^{\uparrow}, \sigma^+) + (M^{\downarrow}, \sigma^-) - (M^{\downarrow}, \sigma^+)/4]$, where $\mu_{\theta_{\mathrm{K}}}$ is the arithmetic mean of $\theta_{\mathrm{K}}$ for the exposed areas. The estimated $\Delta\mu_{\theta_{\mathrm{K}}}$ value was 85 mdeg. The helicity-dependence of magnetization reversal by NIR pulses is more significant than that by XFEL pulses, in contrast to the smaller NIR MCD. In the NIR spectral range, increasing the pulse duration enhances the helicity dependence of the all-optical magnetization reversal[S1]. Furthermore, all-optical magnetization switching is optimized only within a narrow range of fluence[S1]. Therefore, the short pulse duration and wide pulse-energy distribution of the hard XFEL may decrease the helicity dependence of the all-optical magnetization reversal.

**Supplementary Note 4: Determination of the XFEL spot sizes**

The spot sizes of XFEL pulses were determined using a Ce-doped $Y_3Al_5O_{12}$ (YAG: Ce) crystal scintillator with a thickness of 0.3 mm. Figure S3 shows an optical microscopic image of the scintillator irradiated by XFEL pulses at $E$ = 11.569 keV. At the center of the image, a blight spot of x-ray induced optical luminescence can be observed, and the profile corresponds to the dimension of the focused x-ray beam. The $1/e^2$ semi-major and semi-minor axes of the elliptically focal spot were 19.8 and 15.6 μm, respectively. Hence, the area was $S$ = 970 μm$^2$.

**Supplementary Note 5: Estimation of the number of photons per XFEL pulse**

The XFEL intensity was measured using a Si PIN photodiode. The number of photons, $N_{\mathrm{p}}$, per an XFEL pulse was determined using the following equation:

$$N_{\mathrm{p}} = \left[\frac{QW}{e(1 - T_{\mathrm{S_i}})E}\right], \qquad \text{(S2)}$$

where $Q$ is the charge generated in the photodiode element, $W$ is the energy for creating an electron-hole pair in Si, $e$ is the elementary charge, and $T_{\mathrm{S_i}}$ is the x-ray transmittance of the Si photodiode element. We calculated the $N_{\mathrm{p}}$ values for 3000 XFEL pulses using the parameters listed in Table S2. Figure S4 presents the histogram of $N_{\mathrm{p}}$ values. The arithmetic mean and standard deviation are $\mu_{N_{\mathrm{p}}}$= 3.8 $\times 10^{8}$ photons/pulse and $\sigma_{N_{\mathrm{p}}}$= 2.5 $\times 10^{8}$ photons/pulse, respectively. The relatively high $\sigma_{N_{\mathrm{p}}}$ is mainly attributed to the shot-by-shot spectral fluctuations in the self-amplified spontaneous emission process, which are converted to fluctuations in $N_{\mathrm{p}}$ by the monochromator.

## Supplementary Note 6: X-ray energy absorption by the magnetic multilayer

Because the penetration depth of hard x-ray pulses is of the order of 1 μm, most of the pulse energy penetrates the magnetic multilayer (total thickness of 8.5 nm) to the substrate. Because the refractive index is almost equal to 1 in the hard x-ray spectral range, the absorptance of the multilayer can be estimated using the transmittance expressed by $T = e^{-\alpha d}$, where $\alpha$ and $d$ are the linear absorption coefficient and layer thickness, respectively. The calculated values of $T$ and $\alpha$ for each layer at $E$ = 11.569 keV [S11] are listed in Table S3. Subtracting the product of the transmittance of each layer from unity yields the total absorptance of $r_{\mathrm{abs}} = 3.0 \times 10^{-3}$.

## Supplementary Note 7: Comparison of energy absorption between hard x-ray and NIR cases

We evaluated the energy absorbed by the multilayer required for all-optical magnetization reversal using hard x-ray and NIR pulses. The laser fluence is expressed by $F = \mu_{N_{\mathrm{p}}} E/S$, where $\mu_{N_{\mathrm{p}}} E$ is the average pulse energy, and $S$ is the beam spot area. For the XFEL experiment at $E$ = 11.569 keV, $F$ = 72 mJ/cm$^2$ was obtained using $\mu_{N_{\mathrm{p}}} E$ = 0.70 μJ/pulse and $S$ = 970 μm$^2$. The energy absorption per unit volume of the sample is expressed by

$$\rho_{\mathrm{abs}} = \frac{r_{\mathrm{abs}} F}{d_{\mathrm{t}}}, \tag{S3}$$

where $r_{\mathrm{abs}} = 3.0 \times 10^{-3}$ is the total absorptance determined above, and $d_{\mathrm{t}}$ = 8.5 nm is the total thickness of the multilayer. Next, $\rho_{\mathrm{abs}}$(XFEL) = 2.5 $\times 10^{8}$ J/m$^3$ was obtained for the irradiation of an XFEL pulse. In contrast, the energy absorption per unit volume required for the all-optical magnetization reversal at $E$ = 1.55 eV was $\rho_{\mathrm{abs}}$(NIR) = 2.3 $\times 10^{8}$ J/m$^3$, using $F$ = 0.50 mJ/cm$^2$ and $r_{\mathrm{abs}} = P_{\mathrm{a}} / P_{\mathrm{i}}$ = 0.39. The hard x-ray and NIR cases show consistency in energy absorption required for all-optical magnetization reversal. Similar to the NIR case [S12], the all-optical helicity-dependent magnetization reversals in the hard x-ray spectral range may initially require sufficient energy absorption to form completely demagnetized (multi-domain) states, which comprise magnetic domain walls movable by circularly polarized optical pulses.

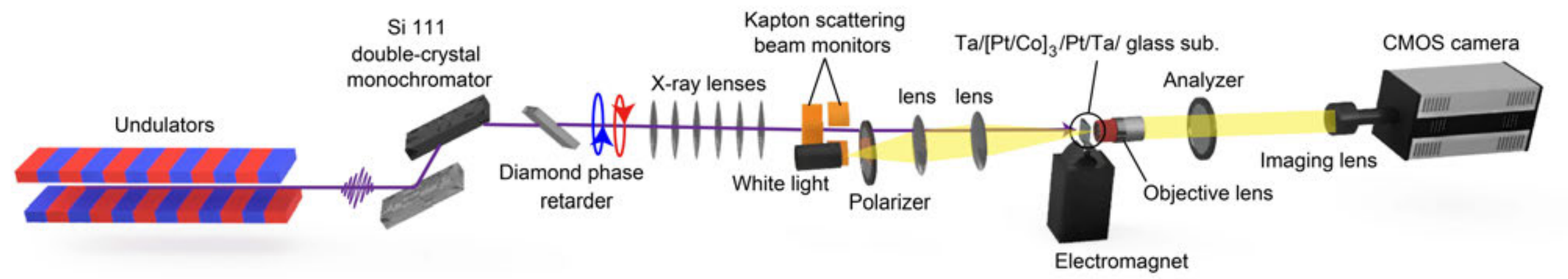

FIG. S1. Schematic of experimental setup for XFEL measurements.

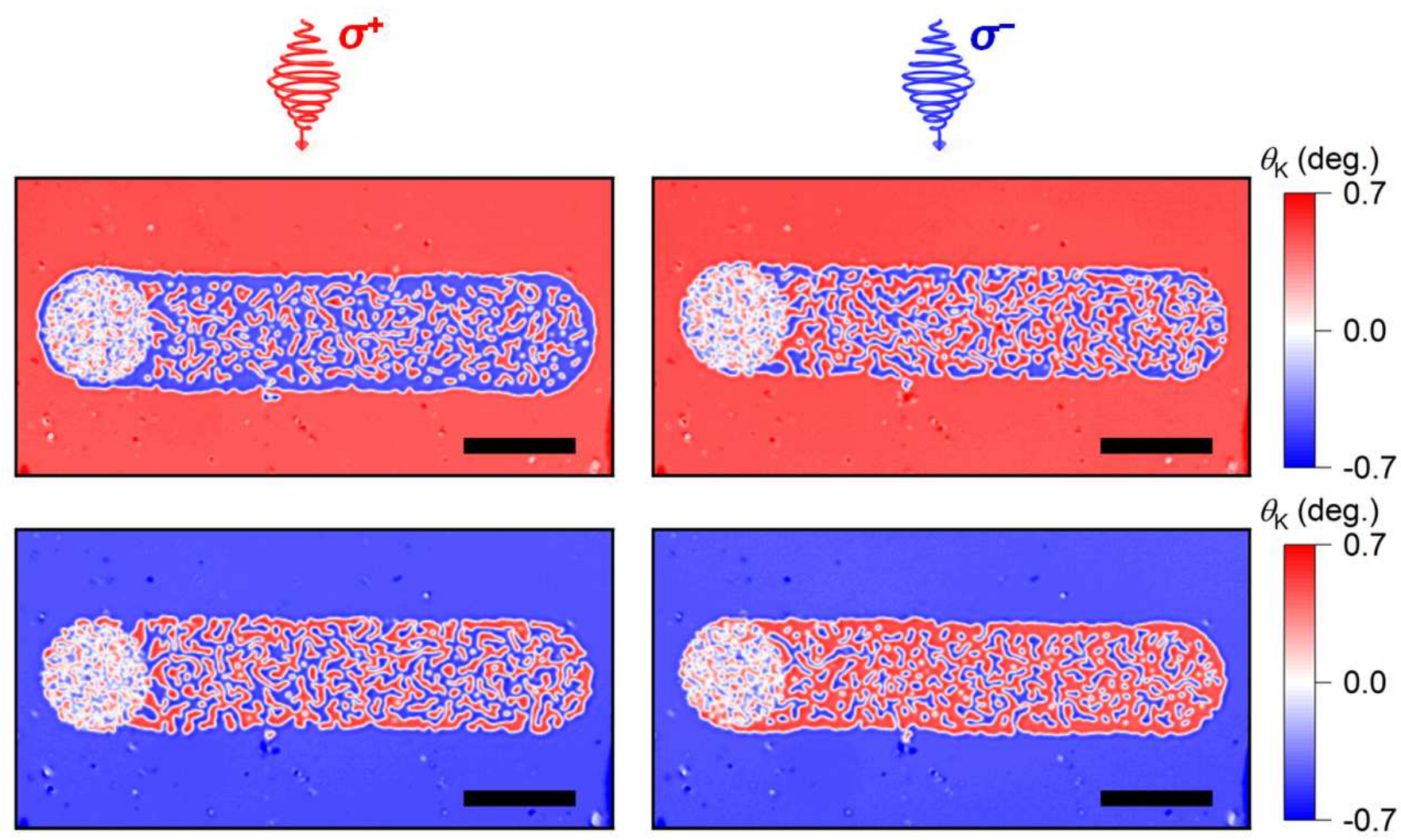

FIG. S2. All-optical magnetization switching by circularly polarized optical pulses at $E$ = 1.55 eV. The red and blue regions indicate up-magnetized and down-magnetized states, respectively. The scale bars correspond to 50 μm.

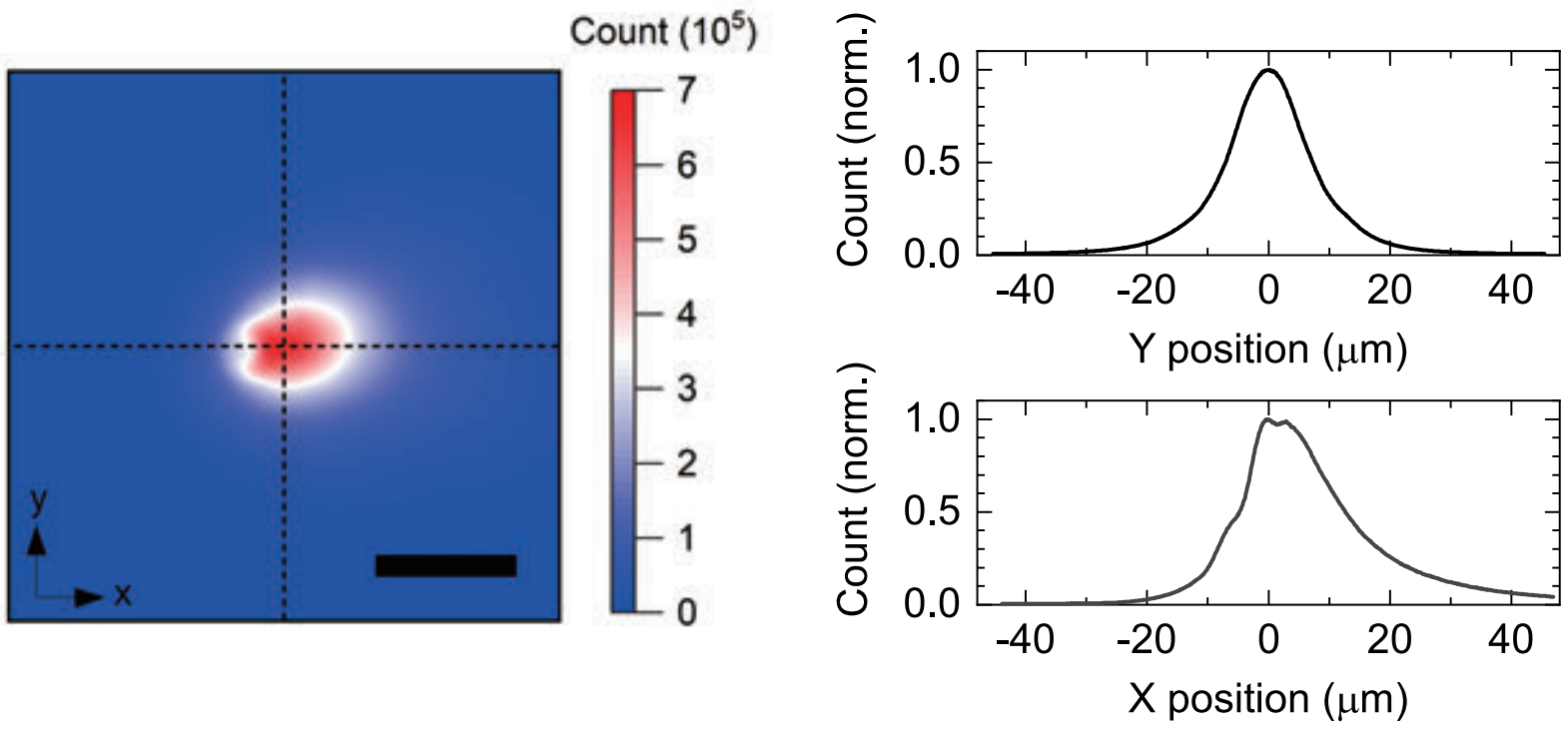

FIG. S3. Focal XFEL beam profile. (Left) Optical microscope image of the X-ray-induced luminescence from a YAG:Ce crystal irradiated by XFEL pulses at $E$ = 11.569 keV. The scale bar corresponds to 20 μm. (Right) Normalized intensity cross-sections along the dashed lines in the left figure.

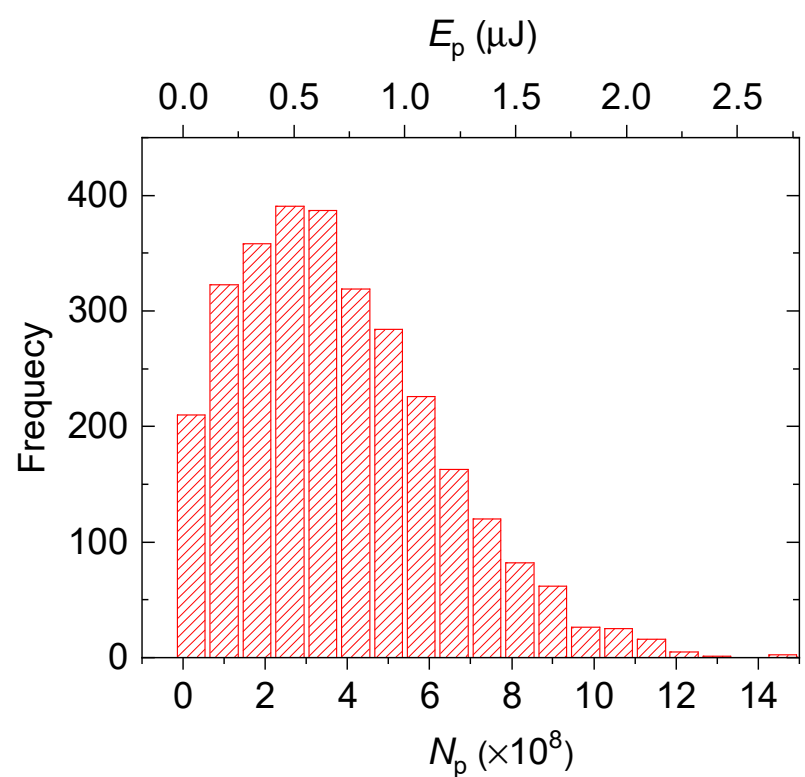

FIG. S4. Distribution of XFEL pulse intensity. A histogram of the number of photons, $N_p$, per XFEL pulse for 3000 shots. The top axis indicates the corresponding pulse energy, $E_p$.

|  | $M^{\uparrow}$ |  | $M^{\downarrow}$ |  |
|---|---|---|---|---|
|  | $\sigma^{+}$ | $\sigma^{-}$ | $\sigma^{+}$ | $\sigma^{-}$ |
| Symbol | Values |  |  |  |
| $P_{r}/P_{i}$ | $0.3467 \pm 0.0006$ | $0.3409 \pm 0.0005$ | $0.3425 \pm 0.0008$ | $0.3451 \pm 0.0006$ |
| $P_{r}/P_{i}$ | $0.2587 \pm 0.0004$ | $0.2666 \pm 0.0005$ | $0.2658 \pm 0.0005$ | $0.2593 \pm 0.0007$ |
| $P_{a}/P_{i}$ | $0.3946 \pm 0.0007$ | $0.3926 \pm 0.0007$ | $0.3917 \pm 0.0009$ | $0.3956 \pm 0.0009$ |
| $\delta_{MCD}$ | $0.007 \pm 0.004$ |  |  |  |

Table S1. Reflectance, transmittance, absorbance, and MCD at $E$ = 1.55 eV.

| Symbol | Units | Value |
|---|---|---|
| $E$ | keV | 11.569 |
| $T_{\mathrm{Si}}$ | | 0.2318 |
| $W$ | eV | 3.66 |
| $e$ | C | $1.602 \times 10^{-19}$ |
| $g$ | nC/V | 500 |
| $T_{\mathrm{air}}$ | | 0.759 |

Table S2. Parameters used for calculating the number of photons per XFEL pulse. The transmittances were calculated using the method described in Ref. S11. The thickness and density ($\rho$) of the Si element are 0.3 mm and 2.33 g/cm$^3$, respectively.

|  |  | Pt | Ta | Co |
|---|---|---|---|---|
| Symbol | Units |  | Values |  |
| $d$ | nm | 4.0 | 3.0 | 1.5 |
| $\alpha$ | $\mu m^{-1}$ | 0.418 | 0.3664 | 0.1097 |
| $\rho$ | $g/cm^3$ | 21.45 | 16.65 | 8.90 |
| $T$ |  | 0.9983 | 0.9989 | 0.9998 |

Table S3. Parameters used for calculating x-ray energy absorption.

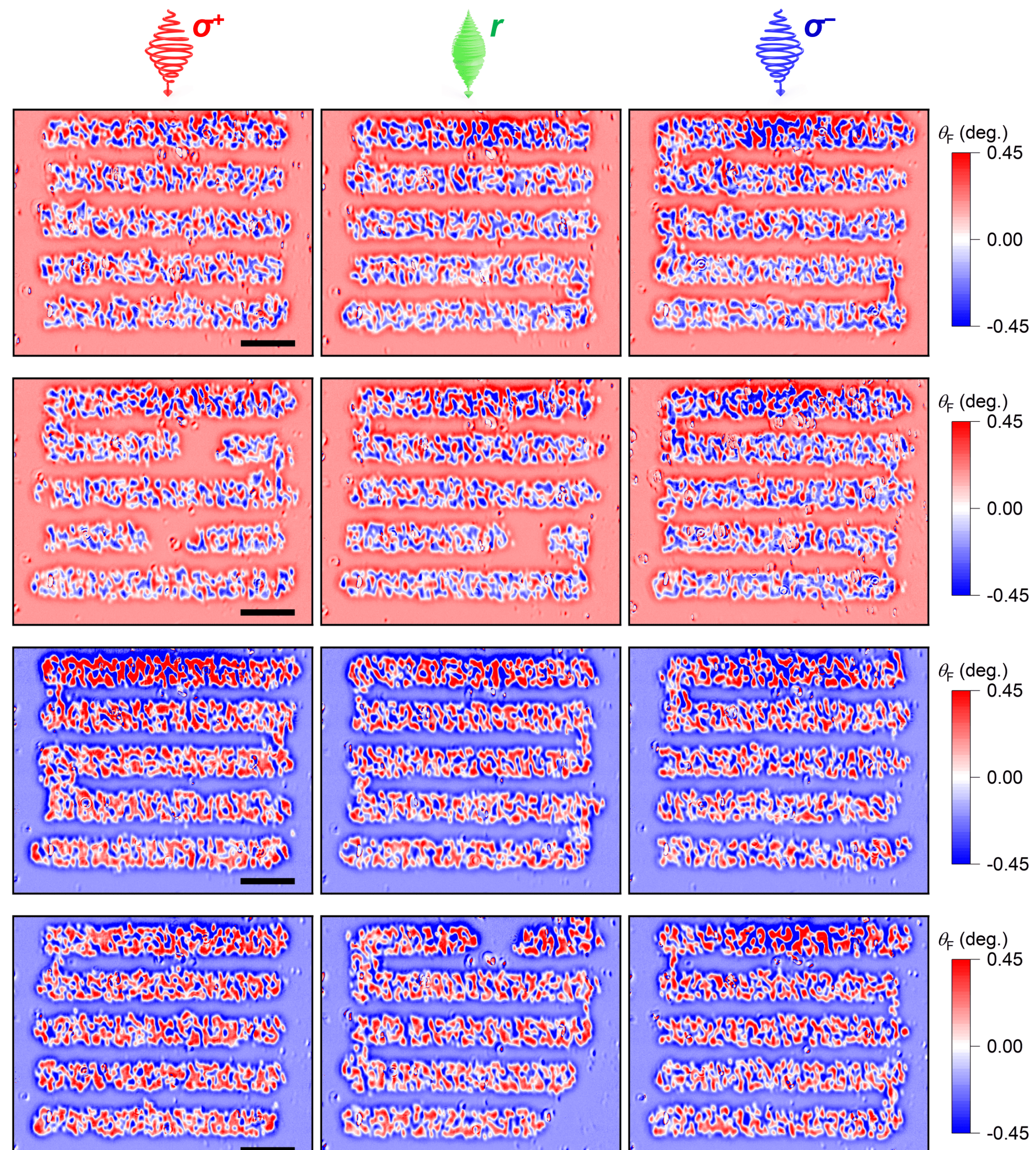

FIG. S5. Magneto-optical images obtained at $E$ = 11.569 keV. The red and blue backgrounds indicate that the original magnetization states were up-magnetized and down-magnetized, respectively. The left, center, and right columns of the images represent the magneto-optical snapshots of the magnetic multilayer irradiated by $\sigma^+$, r, and $\sigma^-$ hard x-ray pulses, respectively. We repeated the line scans twice at each magnetization and polarization condition in the same area. The scale bars correspond to 20 μm. The creation of switched domains occasionally ceased by beam abortions.

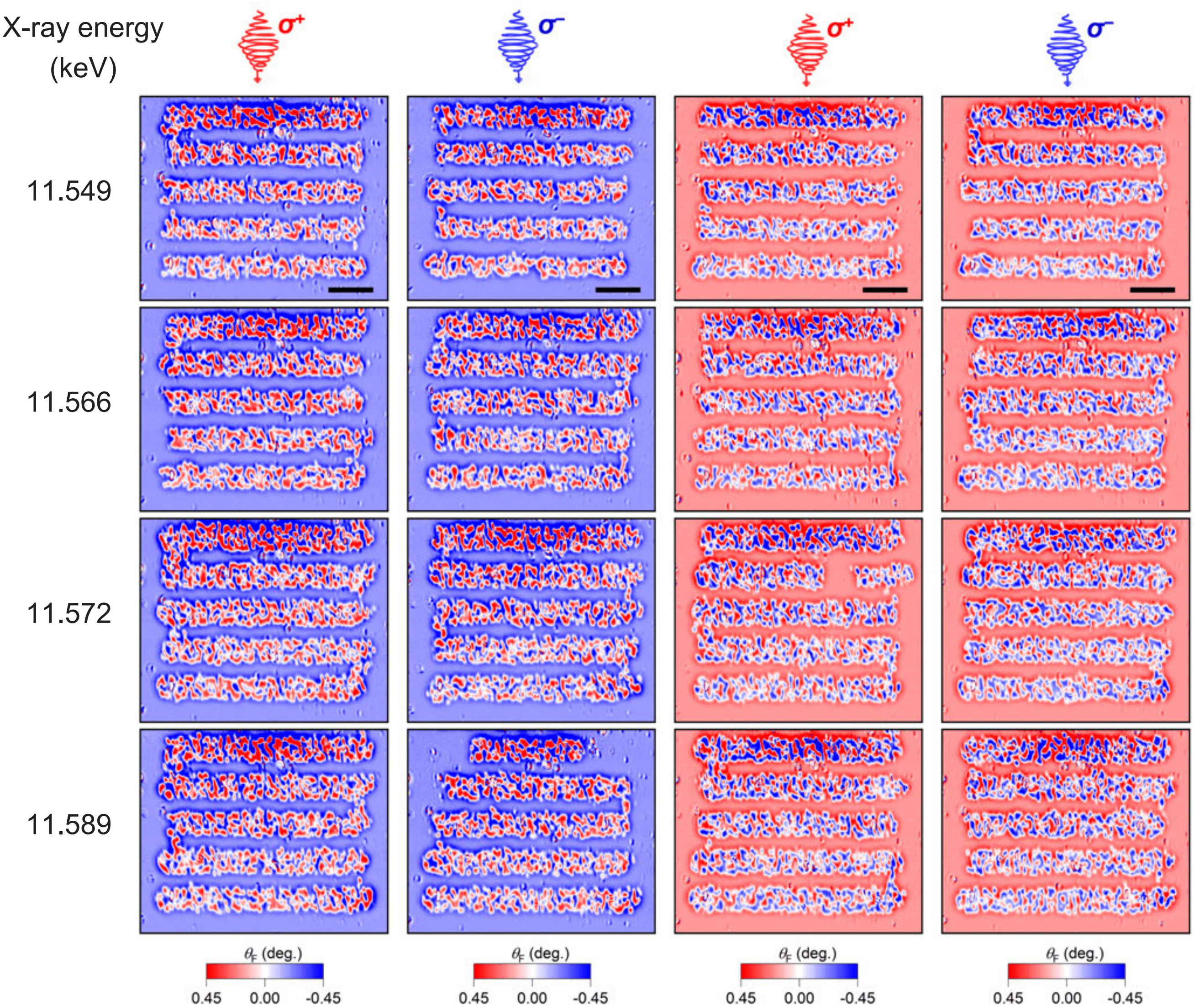

FIG. S6. Magneto-optical images obtained at $E$ = 11.549, 11.566, 11.572, and 11.589 keV. The scale bars correspond to 20 μm. The columns of the images from left to right correspond to cases of ($M^{\downarrow}$, $\sigma^{+}$), ($M^{\downarrow}$, $\sigma^{-}$), ($M^{\uparrow}$, $\sigma^{+}$), and ($M^{\uparrow}$, $\sigma^{-}$), respectively. The rows of the images correspond to the magneto-optical snapshots obtained at $E$ = 11.549, 11.566, 11.572, and 11.589 keV from top to bottom, respectively.

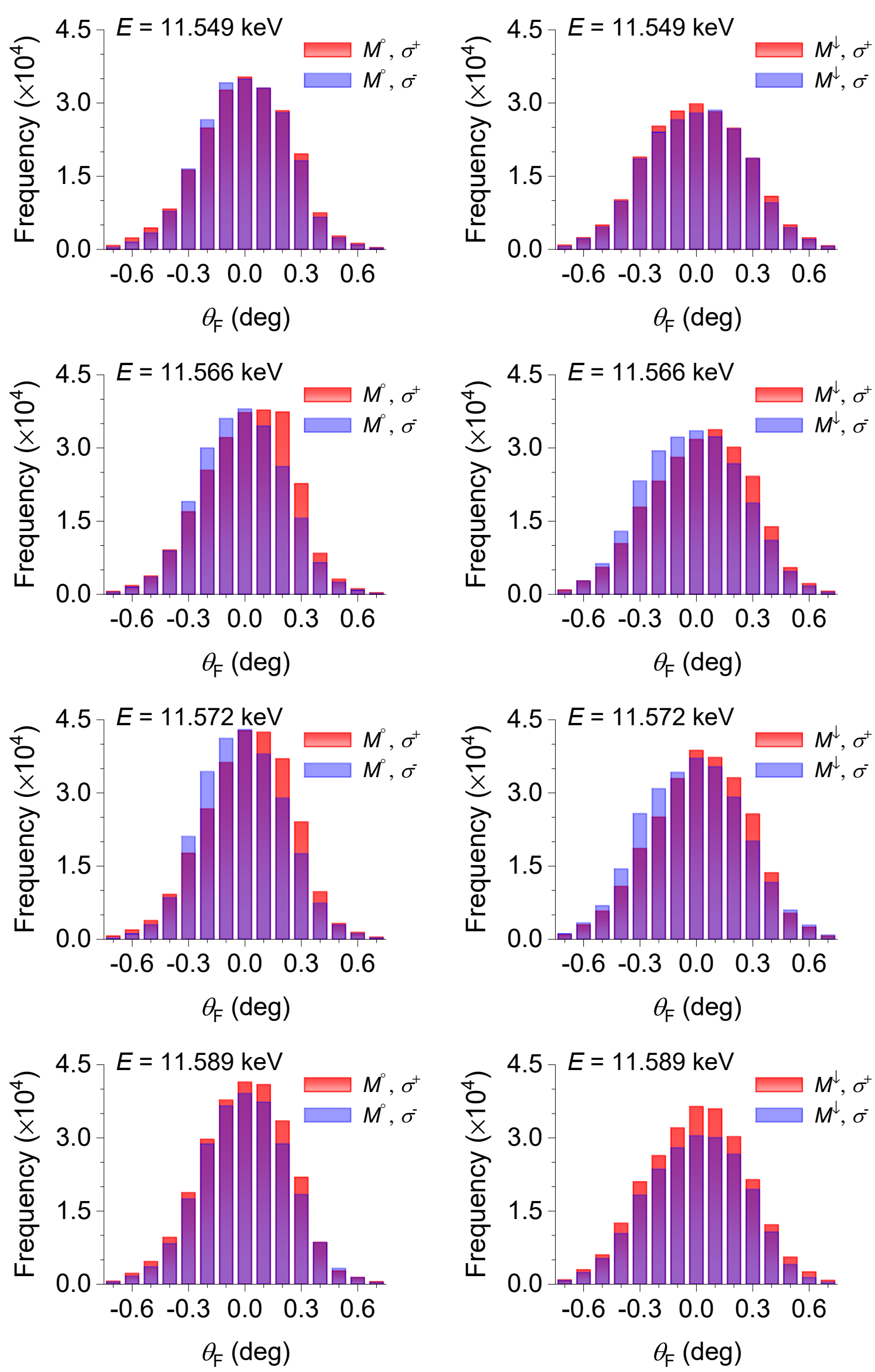

FIG S7. Histograms of $\theta_F$ at $E$ = 11.549, 11.566, 11.572, and 11.589 keV. At 11589 eV, the difference in the total frequencies between the $\sigma^+$ and $\sigma^-$ cases arises mainly from beam abortions and pulse-energy fluctuations. The overall counts increase above the $L_3$ edge because the absorption coefficient of Pt increases.

| | $M^{\uparrow}$ | | $M^{\downarrow}$ | | |
|---|---|---|---|---|---|
| | $\sigma^{+}$ | $\sigma^{-}$ | $\sigma^{+}$ | $\sigma^{-}$ | |
| $E$ (keV) | $\mu_{\theta_F}$ (mdeg) | | | | $\Delta\mu_{\theta_F}$ (mdeg) |
| 11.549 | −1.7 ± 1.2 | −3.4 ± 1.1 | 0.1 ± 1.3 | 1.3 ± 1.3 | −0.3 ± 1.2 |
| 11.566 | 14.8 ± 1.1 | −18.0 ± 1.0 | 20.3 ± 1.2 | −18.1 ± 1.2 | −17.8 ± 1.1 |
| 11.569 | 29.3 ± 1.2 | −27.3 ± 1.1 | 31.0 ± 1.2 | −30.2 ± 1.0 | −29.5 ± 1.2 |
| 11.572 | 15.9 ± 1.2 | −11.5 ± 0.9 | 18.2 ± 1.1 | −15.0 ± 1.1 | −15.1 ± 1.1 |
| 11.589 | 0.3 ± 1.1 | −2.7 ± 1.1 | 2.6 ± 1.1 | −0.7 ± 1.2 | −0.8 ± 1.2 |

Table S5. Arithmetic mean and standard error of $\theta_F$ under each experimental condition. The error of the helicity-dependent part ($\Delta\mu_{\theta_F}$) corresponds to the maximum error.